\documentstyle[12pt]{article}

\newcommand{\Section}[1]{\section{#1}}

\def\sqr#1#2#3#4{{\vcenter{\vskip -#3 pt\hbox{\kern #4 pt\vbox{\hrule height
.#2 pt\hbox{\vrule width .#2 pt height #1 pt\kern #1 pt\vrule width .#2 pt}
\hrule height .#2 pt}\kern #4 pt}}}}

       \def\C{\rm {I\kern-.520em C}}

           \def\be{\begin{equation}}
           \def\bea{\begin{eqnarray}}

               \def\ee{\end{equation}}
               \def\eea{\end{eqnarray}}

                   \def\sl{sl(2)}
                   \def\sll{sl(3)}
                   \def\sn{sl(n)}
                   \def\u{U_q(sl(2))}
                   \def\uu{U_q(sl(3))}
                   \def\un{U_q(sl(n))}
                   \def\unk{U_q\widehat{(sl(n))}}

                       \def\d{\partial}
                       \def\D{D^+}

                           \def\s{\sum}
                           \def\a{\alpha}
                           \def\p{\prod}

                               \def\q{\frac{1}{1+q}}
                               \def\qq{\frac{1}{q-q^{-1}}}

                                   \def\rar{\rightarrow}

\begin{document}

\begin{titlepage}

\begin{flushright}
                                 HEP-TH/9408173
                                 IPM-94-052\\
                                 SUTP/2/73 \\
                                 July 1994
                                             \end{flushright}

\vspace{24pt}
\begin{center}
\begin{large}
             {\bf $\un$ Difference Operator Realization
\footnote{Research partialy supported by Sharif University of Technology}
}
 \end{large}

\

\

\

                          {\bf Azizollah Shafiekhani}
\footnote{e-mail address: ashafie@irearn.bitnet and
ashafie@netware2.ipm.ac.ir}\\
\vspace{6pt} {\it
                 Dept. of Physics\\
                 Sharif University of Technology,\\
                 P.O. Box: 11365-9161, Tehran, Iran\\

\vspace{6pt}
                 and\\

             Institute for Studies in Theoretical Physics and Mathematics\\
             P.O.Box: 19395-5746, Tehran, Iran}\\

\vspace{1.5cm}

\

\

\end{center}
\abstract{

A unified and systematic scheme for construction of differential operator
realization of any irreducible representation of $\sn$ is developed. The
$q$-analogue of this unified scheme is used to construct $q$-difference
operator realization of any irreducible representation of $\un$. Explicit
results for $\u$, $\uu$ and $\un$ are given.

\vfill
}

\end{titlepage}
%%%%%%%%%%%%%%%%%%%%%%%%%
{\Section {Introduction}}

Differential operator realization of Lie algebras are well-known\cite{Z}
and useful in discussing physical systems, both in Quantum Mechanics
\cite{S} and Field Theory\cite{BMP}. There is a well-known correspondence
[3,4] between differential operator realization of Lie algebras[1,3]
and free field representation[5-9] of Kac-Moody algebras[10-11]. A natural
question is whether such a correspondence exists in the case of the deformed
algebras too. More precisely, one may ask if there is any relation between
free field representation[13-18] of quantum affaine Kac-Moody algebras(e.g.
$\unk$)[19-22] and $q$-difference operator realization of
quantum algebras(e.g. $\un$)[23-25].

To address this problem, the first step would be to find a $q$-difference
operator realization of $\un$. There has been an attempt for such a
realization in ref.\cite{FV}, but it is limited to symmetric
representations of $\uu$
\footnote{
          In ref.\cite{FV}, higher dimensional representations have been
          constructed by taking tensor products of fundamental
          representations and the coproduct to define the action of the
          generators on the basis vectors of the algebra. Then $q$-difference
          operator realization of $\uu$ are constructed. By this method only
          symmetric irreducible representation can be obtained. Due to its
          complexity, is not generalizable to higher rank algebras.
          }.
Our aim in this letter is to present, in a unified scheme, a $q$-difference
operator realization of $\un$. One of the new features of this scheme is
that, one is able to construct such a realization, not only for symmetric
representations but for any irreducible representations.

The structure of this letter is as follows: In section 2, we will reproduce
the differential realization of $\sl$, $\sll$ and $\sn$, for any irreducible
representation, by a unified scheme.
In section 3, we will go through $\u$ as a simple example and $\uu$ as a
non-trivial example and finally we will present the $q$-difference operator
realization of $\un$, for any irreducible
representations. \\
\vspace {1.0cm}
%%%%%%%%%%%%%%%%%%%%%%%%%%%%%%%%%
{\Section {Differential operator realization of $sl(n)$}}

The simple case of $sl(2)$ will be considered first in order to explain the
procedure. The well-known $sl(2)$ finite-dimensional representations with one
complex variable will be recovered by a systematic method. Then we shall
take $sl(3)$ as a second example of this procedure and next we will
generalize them for $sl(n)$.

\
\

%%%%%%%%%%%%%%%%%%%%%%%%%%%%%%%%%%

\noindent
{\bf 2.1 $sl(2)$ realization}
\vspace{.50cm}

There are two ways to construct the differential operator representations of
$sl(2)$.  The first method is to start from the fundamental representation
and by tensor multiplying these fundamental irreducible representation,
construct one (two) variable(s) differential operator(s) realization of
higher dimensional representations, which can be found in \cite{S} and
\cite{FV}. In this method one can find realizations of only the symmetric
representation. Here we proceed along a different line in order to find such
realizations for all kinds of representations (note that this difference,
doesn't show itself, for $\sl$.). However, the systematic and
generalizable method is the second one, which we will now describe.
Let $X^\pm$ and $H$ be the Chevalley basis generators of $sl(2)$ with
commutation relations

\be
      [X^+, X^-]=H \hspace{2cm}[H, X^\pm]=\pm2X^\pm
\ee
and let $\mu$ be any arbitrary highest weight,

\be
      X^+{\mid}{\mu}>=0, \hspace{2cm} H{\mid}{\mu}>=2\mu{\mid}{\mu}>.
\ee
Now consider the states

\be
      e^{zX^-}{\mid}{\mu}>
\ee
in the representation space, where $z$ is a complex variable.
The basis vectors  of this vector space are ${\mid}{\mu}>,
X^-{\mid}{\mu}>, ..., (X^-)^{2\mu}{\mid}{\mu}>$.

By the action of $X^-$, $X^+$ and $H$ on (3) and using the above
commutation relations we will have:
$$
      X^-e^{zX^-}{\mid}{\mu}>=\d_z e^{zX^-}{\mid}{\mu}>
$$
\be
      X^+e^{zX^-}{\mid}{\mu}>=(2z\mu-z^2\d_z)e^{zX^-}{\mid}{\mu}>
\ee
$$
    He^{zX^-}{\mid}{\mu}> = (2\mu -2z\d_z)e^{zX^-}{\mid}{\mu}>;\nonumber \\
$$
let define
\be
    J^+=\d_z
\ee
\be
    J^-=2z\mu-z^2\d_z
\ee
\be
    H=2\mu -2z\d_z
\ee
where $\d_z=\frac{\d}{\d z}$. One can consider this, as the representation
of $\sl$ on the sub-space of analytic functions spanned by the monomials,
$\{1,z,z^2,...,z^{2\mu}\}$. We will find that this new operator
representation will satisfy the algebra of (1). The above
realization is a one-complex variable differential operator representation
of $sl(2)$, when $z\in \C$. To see the complications involved in the general
case, we will take $sl(3)$ as a second example.
\

\

%%%%%%%%%%%%%%%%%%%%%%%%%%%%%
\noindent
{\bf 2.2 $sl(3)$ realization}
\vspace{.5cm}

In this case the states in the representation space will be:
\be
    e^{x_1X^-_1}e^{x_{12}X^-_{12}}e^{x_2X^-_2}{\mid}{\mu}>
\ee
where the generators $X^-_i$ $(i=1,2)$ correspond to the simple
roots $\a^i$, $X^-_{12}$ is the  generator corresponds to the non-simple root

$$
    \a_{12}=\a^1+\a^2
$$
and $x_1$, $x_2$ and $x_{12}$ are the complex variables.
In this type of ordering the factor corresponding to the descendents of two
roots lies between them. The defining relations for $\sll$ are as follows:
\be
    [X^\pm_1,X^\pm_2]\equiv\pm X_{12}^\pm
\ee
\be
    [X^\pm_i,X^\pm_{12}]=0
\ee
\be
    [X^+_i,X^-_j]=\delta_{i,j}H_j,\hspace{1cm}i,j=1,2
\ee
\be
    [H_i,X^\pm_j]=\pm a_{ij}H_i
\ee
\be
    [H_i,X_{12}^{\pm}]=\pm X_{12}^{\pm}
\ee
\be
    [H_1,H_2]=0
\ee
where $(a)_{ij}$ is the Cartan matrix. The highest weight $\mu$ is defined such
that:
\be
    X^+_i\mid\mu>=0,\hspace{5mm}H_i\mid\mu>=2\a^i\cdot\mu\mid\mu>,
\ee
where $2\a^i\cdot\mu$ is the Dynkin index(i.e. $\frac{2\a^i\cdot\mu}
{\a^i\cdot\a^i}=m^i$) of the representation, and where we have normalized the
roots to $\a^i\cdot\a^i=1$. By considering the action of
generators $(X^\pm,H)$ on the states given by (8), we find the following
differential operators realization for any irreducible representation with:
\be
    J_{\a^1}^+ = \d_{x_1}
\ee
\be
   J_{\a^1}^- = -x^2_1\d_{x_1} + {x_1}({x_2}\d_{x_2}
  -{x_{12}}\d_{x_{12}}+2\a^1\cdot\mu)-{x_{12}}\d_{x_2}
\ee
\be
   J_{\a^2}^+ = \d_{x_2} + x_1\d_{x_{12}}
\ee
\be
   J_{\a^2}^- = -x_2^2\d_{x_2} + x_{12}\d_{x_1}+2\a^2\cdot\mu{x_2}
\ee
\be
   J_{\a_{12}}^+ = \d_{x_{12}}
\ee
\be
   J_{\a_{12}}^- = -x_{12}^2\d_{x_{12}} - x_1(x_{12}\d_{x_1} -
   {x_2}^2\d_{x_2}+2\a^2\cdot\mu{x_2})-x_2x_{12}\d_{x_2}+2\a_{12}
   \cdot\mu x_{12}
\ee
\be
   H_1=-2{x_1}\d_{x_1} + {x_2}\d_{x_2}-{x_{12}}\d_{x_{12}} +2\a^1\cdot\mu
\ee
\be
   H_2 = x_1\d_{x_1} -2{x_2}\d_{x_2} -{x_{12}}\d_{x_{12}}+2\a^2\cdot\mu
\ee
in agreement with ref.[2]. It is clear that all of the algebra elements can
be constructed by $J^\pm_{\a^1}$ and $J^\pm_{\a^2}$.

The above construction is complete for any representation, with Dynkin
indices $(m,n)$

\begin{center}
$m$ \hspace{14pt}$n$
\vspace{1mm}

\hskip -4pt $\circ$\hskip -1pt---\hskip -2pt---\hskip -1pt$\circ$
\end{center}
\noindent
such that $m,n\not=0$. For any asymmetric representations $(m=0)$,
and symmetric representation $(n=0)$,
the realization must be slightly revised as follows:

For asymmetric $(0,n)$ representations, states in the representation space
will be:
\be
     e^{x_{12}X^-_{12}}e^{x_2X^-_2}{\mid}{\mu}>
\ee
since $X^\pm_1\mid\mu>=0$ and $X^+_2\mid\mu>=0$. After repeating the
same procedure as in (m,n) case one finds:
\be
     J_{\a^1}^+ = -{x_2}\d_{x_{12}}
\ee
\be
     J_{\a^1}^- = -{x_{12}}\d_{x_2}
\ee
\be
     J_{\a^2}^+ = \d_{x_2}
\ee
\be
     J_{\a^2}^- = -{x_2^2}\d_{x_2} + {x_2}(-x_{12}\d_{x_{12}}+2\a^2\cdot\mu)
\ee
\be
     H_1 = x_2\d_{x_2} -x_{12}\d_{x_{12}}
\ee
\be
     H_2 = -2x_2\d_{x_2} -x_{12}\d_{x_{12}}+2{\a^2}.{\mu}.
\ee
For symmetric representations $(m,0)$, states in representation space will be:
\be
     e^{x_1X^-_1}e^{x_{12}X^-_{12}}{\mid}{\mu}>
\ee
and the generators are foud to be:
\be
     J_{\a^1}^+ = \d_{x_1}
\ee
\be
     J_{\a^1}^- = -{x_1^2}\d_{x_1}+x_1(-x_{12}\d_{x_{12}}+2\a^1\cdot\mu)
\ee
\be
     J_{\a^2}^+ = x_1\d_{x_{12}}
\ee
\be
     J_{\a^2}^- = x_{12}\d_{x_1}
\ee
\be
     H_1 = -2x_1\d_{x_1} -x_{12}\d_{x_{12}}+2\a^1\cdot\mu
\ee
\be
     H_2 = x_1\d_{x_1} -x_{12}\d_{x_{12}}.
\ee
So, the minimum number of variables for $(m,n)$, $(0,n)$ and $(m,0)$
representations are 3, 2, 2 respectively. Compairing expressions
(25-30) and (32-37) with (16-23) one finds the following simple
prescription for obtaining the differential realization for the special
$(m,0)$ and $(0,n)$ cases from the general $(m,n)$ case.\\
1- $(m,0)$ case: set $x_1$ equal to zero and replace $\d_{x_1}$ by
$-x^2\d_{x_{12}}$.\\
2- $(0,n)$ case: set $x_2$ and $\d_{x_2}$ equal to zero.

\

\

%%%%%%%%%%%%%%%%%%%%%%%%%%%%
\noindent
{\bf 2.3 $sl(n)$ realization}
\vspace{.5cm}

The Lie algebra $\sn$ defined by the generators $X^\pm$ and $H_i$
$(i=1,...r=n-1)$ and the following relations:
\be
[H_i^+,H^-_j]=0
\ee
\be
     [X_i^+,X^-_j]=\delta_{i,j}H_i
\ee
\be
     [X_i^+,X^-_j]=0, \hspace{.3cm}if\hspace{.3cm}a_{ij}=0
\ee
\be
     [H_i,X^{\pm}_j]=\pm a_{ij}X^\pm_j
\ee
with Serre relations:
\be
\s^n_{k=0}(-1)^k(^n_k)_q(X^\pm_i)^kX^\pm_j(X^\pm_i)^{n-k}=0,\hskip 1.0cm
i{\not=}j
\ee
in order to construct differential realization of this algebra,
take the following states as states in the representation space:
\be
         \p^r_{i=1}\p^r_{j=1}e^{x_{ij}X^-_{ij}}{\mid}{\mu}>
\ee
where $X^-_{ij}$s are the generators correspond to the non-simple roots
$\a_{ij}=\a^i+\a^{i+1}+ \cdot \cdot \cdot +\a^j$
and
$X^\pm_{ii}\equiv X^\pm_i$
corresponds to the simple roots $\a_i$

By similar calculations as in previous cases we obtain the
following realization for $sl(n)$ with any arbitrary highest weight $\mu$ in
the Chevalley basis:
\be
         J^+_{\a^i}=\d_{x_i}+\s^{i-1}_{j=1}x_{j\; i-1}\d_{x_{ji}}
\ee
\be
         J^-_{\a^i}=\s^{i-1}_{j=1}x_{ji}\d_{x_{j\;i-1}}
         -x_i{\big(}-x_i\d_{x_i}+\s^r_{j=i}\s^r_{k=j}\s^k_{l=j}a_{li}x_
         {jk}\d_{x_{jk}}-2\a^i\cdot\mu{\big)}-\s^r_{j=i+1}x_{ij}\d_{x_
         {i+1\;j}}
\ee
\be
         H_{\a^i}=-\s^r_{j=1}\s^r_{k=j}\s^k_{l=j}a_{il}x_{jk}\d_{x_{jk}}
         +2\a^i\cdot\mu
\ee
where $x_{ii}=x_i$. After straightforward, but lengthy calculation
one finds that the above expressions satisfy the $\sn$ algebra.

For any representation, if the $i$th Dynkin index is zero($\a^i\cdot\mu=0$),
we must set $x_i$ equal to zero and replace $\d_{x_i}$ by
\be
        -\s^r_{j=i+1}x_{i+1\;j}\d_{x_{ij}}
\ee
in the above expressions. In the case of several zero adjacent Dynkin indices
$(\a_i\cdot\mu=\a_{i+1}\cdot\mu=\cdot\cdot\cdot=\a_j\cdot\mu=0)$,
one must set all variables $x_{mm'}$ $i\leq m<m'\leq j$ equal to zero and
replace
$\d_{x_{mm'}}$  by
\be
      -\s^r_{k=m'+1}x_{m'+1\;k}\d_{x_{mk}}
\ee
For example, in case of fundamental representation where all Dynkin indices
but first one are zero, we are left with $n-1=r$ variables.

\
\

\vspace{1cm}
%%%%%%%%%%%%%%%%%%%%%%%%%%%
\noindent
{\Section { $\un$}}

Let us first set the notations. Consider $q$-exponential function\cite{MV}:

\be
         e_q^x=\s^\infty_{n=0}\frac{(1-q)^n x^n}{(1-q)(1-q^2)...(1-q^n)}
\ee
and the $q$-difference operator
\be
         D^\pm_xf(x)=\frac{f(q^{\pm1} x)-f(x)}{(q^{\pm1}-1)x}=
         \frac{1}{(q^{\pm1}-1)x}(M^{\pm1}_x -1)f(x)
\ee
where $M_x$ is a translation operator, defined by $M_x^af(x)=f(q^ax)$.

\
\

%%%%%%%%%%%%%%%%%%%%%%%%%%%
\noindent
{\bf 3.1 $\u$ }\cite{MO} {\bf and $\uu$ realization}
\vspace{.5cm}

Just as in the case of ordinary $sl(2)$, we take the states
\be
e^{zX^-}_q\mid\mu>
\ee
in the representation space. The algebra relations are:

\be
[X^+, X^-]=\frac{q^H-q^{-H}}{q-q^{-1}} \hspace{2cm}[H, X^\pm]=\pm2X^\pm
\ee
with the same procedure as of section 2.1 one will find the following
$q$-difference operator realization for any arbitrary highest weight
representation:
\be
J^+=\D_z
\ee
\be
J^-=\qq z(M^{-2}_z q^{2\mu}-q^{-2\mu})
\ee
\be
H=-2z\d_z+2\mu
\ee
in agrement with \cite{FV}.

%%%%%%%%%%%%%%%%%%%%%%%%
\noindent
\vspace{0.4cm}

For $\uu$ realization we take
\be
e_q^{x_1X^-_1}e^{x_{12}X^-_{12}}_qe^{x_2X^-_2}_q\mid\mu>
\ee
as the states in the representation space. This type of ordering is well-known
in quantum algebra, and called "normal" in [29-30]. The algebra relations are:

$$[X_i^+,X^-_j]=\delta_{i,j}\frac{q^{H_i}-q^{-H_i}}{q-q^{-1}}\equiv
\delta_{i,j}[H_i]_q$$

$$[X_1^{\pm},X^{\pm}_2]_{q^{\pm1}}=q^{\pm1/2} X^\pm_1 X^\pm_2-q^{\mp1/2}
X^\pm_2X^\pm_1 =\pm X^\pm_{12}$$

\be
[X_i^{\pm},X^{\pm}_{12}]=q^{\mp1/2} X^\pm_1 X^\pm_{12}-q^{\pm1/2}
X^\pm_{12}X^\pm_1 =0
\ee

$$[H_i,X^{\pm}_{j k}]=\pm{\big(}\s^k_{l=j}a_{il}{\big)}X^\pm_{j k}$$

$$[H_i,H_j]=0.$$
Now by the action of $X_i^\pm$ and $H_i$ ($i,j=1, 2$) on the above states, we
will have the following $q$-difference operator realization for $\uu$:
\be
J_{\a^1}^+ = D_{x_1}
\ee
\bea
J_{\a^1}^-&=&\frac{1}{q+1}{\bigg[}\qq(1+M^{+1}_{x_1})x_1
(M^{-2}_{x_1} M^{-1}_{x_{12}} M^{+1}_{x_2} q^{2\a^1\cdot\mu}-M^{+1}_{x_{12}}
M^{-1}_{x_2}q^{-2\a^1\cdot\mu})\nonumber \\
&-&q^{-1/2}(1+ M^{+1}_{x_{12}})x_{12}M^{+1}_{x_2}q^
{-2\a^1\cdot\mu}\D_{x_2}{\bigg]}
\eea
\be
J_{\a^2}^+=\frac{q^{1/2}}{q+1}M^{-1}_{x_1}(1+M^{+1}_{x_1})x_1\D_{x_{12}}
+M^{-1}_{x_1}M^{+1}_{x_{12}}\D_{x_2}
\ee

\bea
J_{\a^2}^-&=&\q {\bigg[}q^{1/2}(1+M^{+1}_{x_{12}})x_{12}M^{-1}_{x_{12}}
M^{-2}_{x_2}q^{2\a^2\cdot\mu}\D_{x_1} \nonumber \\
&+&\qq(1+M^{+1}_{x_2})x_2(M^{-2}_{x_2}q^{2\a^2\cdot
\mu}-q^{-2\a^2\cdot\mu}){\bigg]}
\eea
\be
H_1 = -2x_1\d_{x_l} - x_{12}\d_{x_{12}} +x_2\d_{x_2}+2\a^1\cdot\mu
\ee
\be
H_2 = x_1\d_{x_l} - x_{12}\d_{x_{12}} -2x_2\d_{x_2}+2\a^2\cdot\mu.
\ee

As before the above elements of the algebra will satisfy the corresponding
algebra for any arbitrary representation with non-zero Dynkin indices.
For asymmetric case where the first Dynkin index is zero, we should set
$x_1$ equal to zero and $\D_{x_1}$ must be replace by

$$-\frac{q^{1/2}}{1+q}M^{-1}_{x_{12}}(1+M^{+1}_{x_2})x_2\D_{x_{12}}$$
in the algebra elements. The final result for asymmetric
representations $(0,n)$ are as follows:
\be
J_{\a^1}^+ =-\frac{q^{1/2}}{1+q}M^{-1}_{x_{12}}(1+M^{+1}_{x_2})x_2\D_{x_{12}}
\ee
\be
J_{\a^1}^-=-\frac{q^{3/2}}{1+q}M^{-2}_{x_{12}}(1+M^{+1}_{x_{12}})x_{12}
M^{-1}_{x_2}\D_{x_2}
\ee
\be
J_{\a^2}^+=M^{+1}_{x_{12}}\D_{x_2}
\ee
\bea
J_{\a^2}^-&=&\q{\Bigg[}-q^{-1}(1+M^{+1}_{x_{12}})x_{12}
M^{-2}_{x_{12}}M^{-2}_{x_2}(1+M^{+1}_{x_2})x_2 q^{2\a^2\cdot\mu}
\D_{x_{12}}\nonumber \\
&+&\qq{\big(}1+M^{+1}_{x_2}{\big)}x_2{\Big(}M^{-2}_{x_2}q^{{2\a^2}\cdot\mu}-
q^{-2\a^2\cdot\mu}{\Big)}{\Bigg]}
\eea
\be
H_1 = - x_{12}\d_{x_{12}} +x_2\d_{x_2}
\ee
\be
H_2 = - x_{12}\d_{x_{12}} -2x_2\d_{x_2}+2\a^2\cdot\mu
\ee
Similarly, where the second Dynkin index is zero, we should set
$x_2$ equal to zero, and replace $\D_{x_2}$ by zero. So, the final result
for symmetric $(m,0)$ representations will be as follows:
\be
J_{\a^1}^+ =\D_{x_1}
\ee
\be
J_{\a^1}^-=\q\qq(1+M^{+1}_{x_1})x_1{\big(}M^{-2}_{x_1}M^{-1}_{x_{12}}
q^{2\a^1\cdot\mu}-M^{+1}_{x_{12}}q^{-2\a^1\cdot\mu}{\big)}
\ee
\be
J^+_{\a^2}=\frac{q^{1/2}}{1+q}(1+M^{-1}_{x_1})x_1\D_{x_{12}}
\ee
\be
J^-_{\a^2}=\frac{q^{1/2}}{q+1}M^{-1}_{x_12}(1+M^{+12}_{x_1})x_1\D_{x_{1}}
\ee
\be
H_1 = -2x_1\d_{x_1}-x_{12}\d_{x_{12}}+2\a^1\cdot\mu
\ee
\be
H_2 = x_1\d_{x_1}-x_{12}\d_{x_{12}}
\ee

\
\

%%%%%%%%%%%%%%%%%%%
\noindent
{\bf 3.2 $\un$ realization}
\vspace{.5cm}

For the general case $\un$ we take the following states in representation
space:
\be
\p^r_{i=1}\p^r_{j=1}e_q^{x_{ij}X^-_{ij}}{\mid}{\mu}>
\ee
The algebra of $\un$ in the Chevalley basis can be summerized as
follows[23-24]:
\be
[H_i,H_j]=0
\ee
\be
[H_i,X^{\pm}_j]={\pm}a_{ij}X^{\pm}_j
\ee
\be
[X_i^+,X^-_j]=\delta_{i,j}\frac{q^{H_i}-q^{-H_i}}{q-q^{-1}}
\ee
\be
[X_i^\pm,X^\pm_j]=0,\hskip 1.5cm if \hskip 0.75cm a_{ij}=0
\ee
\noindent
with Serre relations,
\be
\s^n_{k=0}(-1)^k(^n_k)_q(X^\pm_i)^kX^\pm_j(X^\pm_i)^{n-k}=0,\hskip 1.5cm
i{\not=}j
\ee
with $a_{ij}=\frac{2(\a^i\cdot\a^j)}{(\a^i\cdot\a^i)}$, Cartan matrix,
$n=1-a_{ij}$ and
\be
(^n_k)_q={\frac{[n]_q!}{[k]_q![n-k]_q!}},\hskip
1.5cm{[m]_q!=[m]_q[m-1]_q...[1]_q},
\hskip 1.5cm{[m]_q=\frac{q^m-q^{-m}}{q-q^{-1}}}
\ee

By the same procedure, difference operator representation of algebra elements
in Chevalley basis are as follows:
\bea
J^+_{\a^i}&=&{\Bigg(}\delta_{i,1}+\p^{i-1}_{j=1}\p^r_{k=j}
M^{\s_{l=j}^k a_{li}}_{x_{jk}}{\Bigg)}\D_{x_i}\nonumber \\
&+&\frac{q^{1/2}}{1+q}\s^{i-1}_{j=1}\p_{k=1}^j\p_{l=k}^{i-1}
M^{\s_{m=k-1}^i a_{mi}}_{x_{k-1\; i}} M^{\s_{n=k}^l a_{ni}}_{x_{kl}}
(1+M^{+1}_{x_{j\; i-1}})x_{j\; i-1} \D_{x_{ji}}
\eea
\bea
J^-_{\a^i}&=&\q\Bigg{\lbrack}
q^{1/2}\s_{j=1}^{i-1}(1+M^{+1}_{x_{ji}})x_{ji}\prod_{k=i}^r
M^{- \sum_{l=j}^k a_{li}}_{x_{jk}} \prod_{m=j+1}^r\prod_{n=m}^r
M^{-{\sum}_{p=m}^na_{pi}}_{x_{mn}}q^{2{\a^i}\cdot\mu}\D_{x_{j\;i-1}}\nonumber
\\
&+&\qq (1+M^{+1}_{x_i})x_i {\Bigg (}\p^r_{j=i}\p^r_{k=j}
M^{-{\sum}_{l=j}^ka_{li}}_{x_{jk}}q^{2{\a^i}\cdot\mu} - M^{-2}_{x_{ii}}
\p^r_{j=i}\p^r_{k=j}M^{{\sum}_{l=j}^ka_{li}}_{x_{jk}}q^{-2{
\a^i}\cdot\mu}{\Bigg)} \nonumber
\eea
\be
-q^{-1/2}\s^r_{j=i+1}(1+M^{+1}_{x_{ij}})x_{ij}M^{-1}_{x_{ij}}\p_{k=j}^r
M^{\s_{l=i}^k a_{il}}_{x_{ik}} \p_{m=i+1}^r\p_{n=m}^r M^{\s_{p=m}^n
a_{ip}}_{x_{mn}}\p_{s=i+1}^{j-1}M^{+(1-\delta_{i+1,j})}_{x_{i+1\;s}}
q^{-2{\a^i}\cdot\mu}\D_{x_{i+1\; j}}{\Bigg\rbrack}
\ee

\be
H_{\a^i}=-\s^r_{j=1}\s^r_{k=j}\s^k_{l=j}a_{il}x_{jk}\d_{x_{jk}}+2\a^i\cdot\mu
\ee
\noindent
with $i=1,...r$, $r=n-1$ rank of the group, and $M_{x_{i0}}=1$.

For the representation with some zero Dynkin indices as stated before we must
set corresponding variables to zero and corresponding difference opreator
should be replaced as follows:
\be
\D_{x_{mm'}}\rar-\frac{q^{1/2}}{q+1}\s^r_{i=m'+1}\p^i_{j=m'+1}M^{-1}_{x_{mj}}
\p_{k=m+1}^{m'}M^{+1}_{x_{ki}}\p_{l=m'+1}^{i-1}M^{-(\delta_{i,m'+1}-1)}_
{x_{m'+1\;l}}(1+M^{+1}_{x_{m'+1\;i}})x_{m'+1\;i}\D_{x_{m'i}}
\ee
where $x_{mm'}$ corresponds to
$\a_{mm'}=\a^m+\a^{m+1}+\cdot\cdot\cdot+\a^{m'-1}+\a^{m'}\in\Sigma^+$ and
$\a^m\cdot\mu,\cdot\cdot\cdot,\a^{m'}\cdot\mu=0$. Note that
for any fundamental representation the number of variables is $n-1=r$.

In the limit of $q\rar 1$ all the relations for $\un$ will go to the ordinary
$sl(n)$. To be more specific, when $q\rar 1$
\be
\D_{x_{ij}}\rar\d_{x_{ij}}
\ee
and
\be
M^a_{x_{ij}}\rar 1
\ee
equations (83-85) will be exactly the same as the equations (44-46).

We have explicitly checked that the generators defined in eq. (83-85)
satify the algebra given in (77-81) for $n$ up to 6. We have checked the
following relations:
\be
[J^+_{\a^i},J^-_{\a^i}]=\qq(q^H_i-q^{-H_i})
\ee
\be
[H_i,J^\pm_{\a^i}]=\pm2J^\pm_{\a^i}
\ee
\be
[J^\pm_{\a^{i-2}},J^\pm_{\a^i}]=0
\ee
are satisfied for all $n$.

\
\

%%%%%%%%%%%%%%%%%%%%%%
\noindent
{\bf Acknowledgement:}

I am greatful to Prof. A. Morozov for proposing this problem and his
useful advice. I would like also to thank Prof. F. Ardalan for his useful
guidence and useful discussions.

\newpage

%%%%%%%%%%%%%%%%%%%%%%%%%%%%%%%%%%

\end{document}